\documentclass[
showpacs,
floatfix,
aps,
prl,
twocolumn,
superscriptaddress,
amssymb,
]{revtex4-1}

\usepackage{times}
\usepackage{amssymb}
\usepackage{latexsym}
\usepackage[dvips]{graphicx}
\usepackage{amsmath}

\usepackage{graphicx}
\usepackage{dcolumn}
\usepackage{amsfonts}
\usepackage{bm}
\usepackage{epsfig}

\newcommand{\be}{\begin{equation}}
\newcommand{\ee}{\end{equation}}
\newcommand{\bea}{\begin{eqnarray}}
\newcommand{\eea}{\end{eqnarray}}

\def\rxx{R_{xx}}
\def\rxy{R_{xy}}

\def\mcf{m^{\star}}
\def\tqcf{\tau_{\rm q}^{\star}}

\def\occf{\omega_{\rm c}^{\star}}
\def\nucf{\nu^\star}
\def\efcf{E_F^\star}
\def\ncf{n^\star}
\def\occf{\omega_{\rm c}^\star}
\def\gcf{\Gamma^\star}
\def\bcf{B^\star}
\def\ec{E_C}
\def\fq{\phi_0}

\newcommand{\rfig}[1]{Fig.\,\ref{#1}}

\begin{document}
\title{Spinless composite fermions in an ultra-high quality strained Ge quantum well}
\author{Q.~Shi}
\affiliation{School of Physics and Astronomy, University of Minnesota, Minneapolis, Minnesota 55455, USA}
\author{M.~A.~Zudov}
\email[\vspace{-0.2in} Corresponding author: ]{zudov@physics.umn.edu}
\affiliation{School of Physics and Astronomy, University of Minnesota, Minneapolis, Minnesota 55455, USA}
\author{C.~Morrison}
\affiliation{Department of Physics, University of Warwick, Coventry, CV4 7AL, United Kingdom}
\author{M.~Myronov}
\affiliation{Department of Physics, University of Warwick, Coventry, CV4 7AL, United Kingdom}

\begin{abstract}
We report on an observation of a fractional quantum Hall effect in an ultra-high quality two-dimensional hole gas hosted in a strained Ge quantum well.
The Hall resistance reveals precisely quantized plateaus and vanishing longitudinal resistance at filling factors $\nu = 2/3, 4/3$ and $5/3$.
From the temperature dependence around $\nu = 3/2$ we obtain the composite fermion mass of $\mcf \approx 0.4\,m_e$, where $m_e$ is the mass of a free electron.
Owing to large Zeeman energy, all observed states are spin-polarized and can be described in terms of spinless composite fermions.
\end{abstract}
\maketitle

Since its discovery in two-dimensional electron gas hosted in GaAs/AlGaAs more than three decades ago \citep{tsui:1982}, fractional quantum Hall (FQH) effect has been realized only in a select few semiconductor materials \citep{nelson:1992,poortere:2002,lai:2004,piot:2010,kott:2014,betthausen:2014},
graphene \citep{du:2009,bolotin:2009}, and an oxide \citep{tsukazaki:2010}.
The FQH effect can be conveniently viewed as an integer quantum Hall effect \citep{klitzing:1980} of composite fermions, which perform cyclotron motion in a reduced, effective magnetic field $\bcf$ \citep{jain:1989,jain:2007,jain:2015}. 
This motion leads to formation of composite fermion Landau levels, 
termed $\Lambda$-levels, 
which are separated by energy $\hbar\occf = \hbar e \bcf/\mcf$, where $\mcf$ is the composite fermion effective mass. 
However, with an exception of CdTe, in all of the above systems FQH states are strongly influenced by spin and/or valley degrees of freedom, due to comparable energy scales. 
Since these degrees of freedom carry over to composite fermions, the $\Lambda$-level spectrum becomes much more complex and certain FQH states are either very weak or not observed at all. 

A two-dimensional hole gas (2DHG) in a strained Ge quantum well is a single-valley, single-band material system, with out-of-plane component of the Land${\rm\acute{e}}$ $g$-factor which is an order of magnitude larger than for electrons in GaAs \cite{winkler:1996,winkler:2003}.
As such, strained Ge appears to be one of the simplest systems to investigate FQH physics; owing to large Zeeman energy, the FQH effect should originate entirely from the orbital motion of composite fermions, even in the first excited Landau level. 
In addition, Ge is interesting in some other aspects.
First, due to its diamond crystal structure, the Dresselhaus spin-orbit coupling is absent, while the Rashba spin-orbit parameter is comparable to that in GaAs \cite{moriya:2014,morrison:2014}, making Ge a unique material for spintronic applications.
Second, it has been recently found that in tilted magnetic fields Ge exhibits strong transport anisotropy whose underlying mechanism is not yet understood \citep{shi:2015a}.
Finally, Ge can add new functionalities to Si-based devices and sometimes is even viewed as a candidate for non-Si-based semiconductor technologies \citep{pillarisetty:2011}.

While it is still not entirely clear what material parameters determine the quality of the FQH effect \citep{dassarma:2014}, it is well established that sufficiently high carrier mobility is a prerequisite for its observation. 
Indeed, even though strained Ge has been successfully used in studies of quantum Hall liquid - insulator transitions \cite{song:1997,hilke:1997,hilke:2000}, a rather low hole mobility ($\mu < 10^5$ cm$^2$/Vs) precluded observation of the FQH effect in this material system.

In this Rapid Communication we report on a quantum transport measurements in an extremely high-quality ($\mu > 10^6$ cm$^2$/Vs) 2DHG hosted in a strained Ge quantum well \cite{dobbie:2012,zudov:2014,shi:2014b}.
The Hall resistance reveals plateaus at filling factors $\nu = 2/3, 4/3$ and $5/3$ which are quantized precisely at $h/e^2\nu$.
Analysis of the temperature dependence of the Shubnikov-de Hass oscillations of composite fermions around filling factor $\nu = 3/2$ yields the composite fermion mass of $\mcf \approx 0.4\,m_e$.
This value is in good agreement with the theory: at $\nu =5/3$, it translates to the composite fermion cyclotron gap of $\hbar\occf \approx 0.07 e^2/4\pi\epsilon_0\epsilon \ell$, where $\ell = \sqrt{h/eB}$ is the magnetic length, $\epsilon =16$ is the dielectric constant of Ge, and $\epsilon_0$ is the vacuum permittivity. 
Owing to large Zeeman energy, all observed FQH states are spin-polarized and thus can be described by composite fermions with only orbital degree of freedom.

Our sample is a $\approx5\times5$ mm square fabricated from a fully strained, $\approx 17$ nm-wide Ge quantum well grown by reduced pressure chemical vapour deposition on a relaxed Si$_{0.16}$Ge$_{0.84}$/Ge/Si(001) virtual substrate \cite{dobbie:2012,morrison:2014,myronov:2014,myronov:2015}
Holes are supplied by a 12 nm-wide B-doped layer separated from the interface by a 30 nm-wide undoped Si$_{0.16}$Ge$_{0.84}$ spacer. 
At $T$ = 0.3 K, our 2DHG has density $p \approx 2.9 \times 10^{11}$ cm$^{-2}$ and mobility $\mu \approx 1.3 \times 10^6$ cm$^2$/Vs. 
The longitudinal resistance ($\rxx$) and the Hall resistance ($\rxy$) were measured by a low-frequency (a few hertz) lock-in technique in sweeping perpendicular magnetic fields up to $B = 18$ T and temperatures down to $T \approx 0.3$ K.

\begin{figure}[t]
\includegraphics[width=\linewidth]{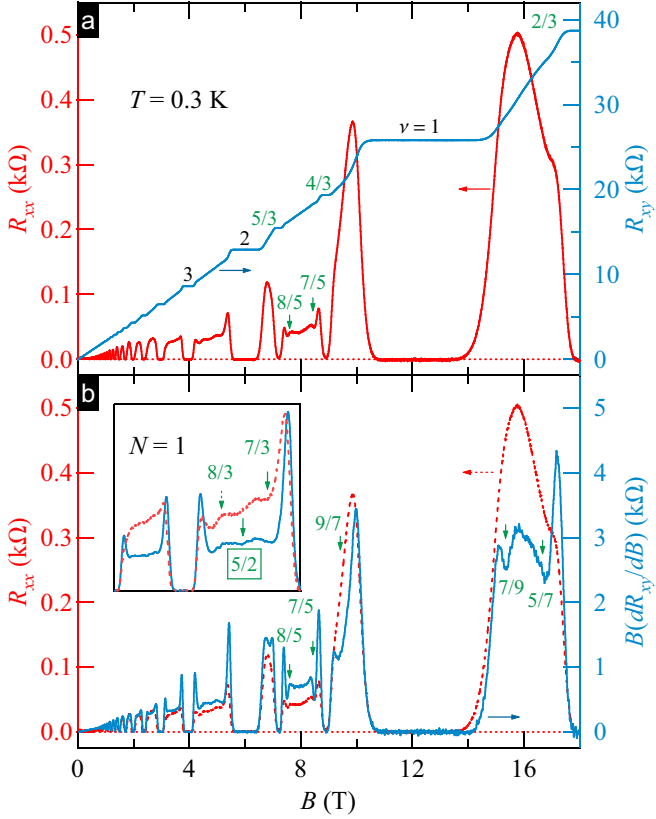}
\vspace{-0.25 in}
\caption{(Color online)
(a) $\rxx$ (left axis) and $\rxy$ (right axis) versus $B$. 
(b) $\rxx$ (left axis, dotted line) and $B\cdot(d\rxy/dB)$ (right axis, solid line) versus $B$.
Integers and fractions next to the traces mark filling factors.
Inset shows $\rxx(B)$ (dotted line) and $B\cdot d\rxy(B)/dB$ (solid line) in the $N=1$ Landau level.
Vertical arrows are drawn at corresponding $\nu$, as marked.
}
\vspace{-0.2 in}
\label{fig1}
\end{figure}

In \rfig{fig1}(a) we present $\rxx$ (left axis) and $\rxy$ (right axis) versus magnetic field $B$ measured at base temperature $T \approx 0.3$ K. 
Vertical arrows are drawn at corresponding $\nu$, as marked.
In addition to integer quantum Hall effect, the $\rxx$ data clearly reveal deep minima in the vicinity of filling factors $\nu=5/3$ and $4/3$, while $\rxy$ shows fully quantized Hall plateaus.
These fractions are the primary states of the series $\nu = 2 - \nucf/(2\nucf \pm 1$), corresponding to composite fermion filling factors $\nucf = 1$ and $\nucf = 2$, respectively.
Closer inspection of the data also reveals weak $\rxx$ minima at $\nu = 8/5$ and $\nu = 7/5$.
In addition, we also observe a fully developed FQH state at $\nu = 2/3$, corresponding to $\nucf = 2$ of the $\nu = \nucf/(2\nucf-1)$ series.

We next examine the applicability of the empirical ``resistance rule'' \citep{chang:1985,pan:2005} which states that $\rxx$ and $\rxy$ are related as $\rxx \sim B\cdot d\rxy/dB$. 
In \rfig{fig1}(b) we present $\rxx (B)$ (left axis, dotted line) and $B\cdot d\rxy(B)/dB$ (right axis, solid line), calculated from the $\rxy(B)$ data shown in \rfig{fig1}(a).
Direct comparison reveals excellent agreement between the two quantities over the entire magnetic field range.
In addition, $B\cdot d\rxy(B)/dB$ reveals strong minima at $\nu = 8/5$ and $\nu = 7/5$, as well as dips near $\nu = 5/7, 7/9$ and $9/7$, indicating developing FQH states at these filling factors.
The inset shows the zoomed-in version of same data in the vicinity of $\nu = 5/2$.
As illustrated by the arrows, very weak minima can be see near $\nu = 5/2$ and $\nu = 7/3$, while $\nu = 8/3$ corresponds to the $\rxx$ maximum.

\begin{figure}[t]
\includegraphics[width=\linewidth]{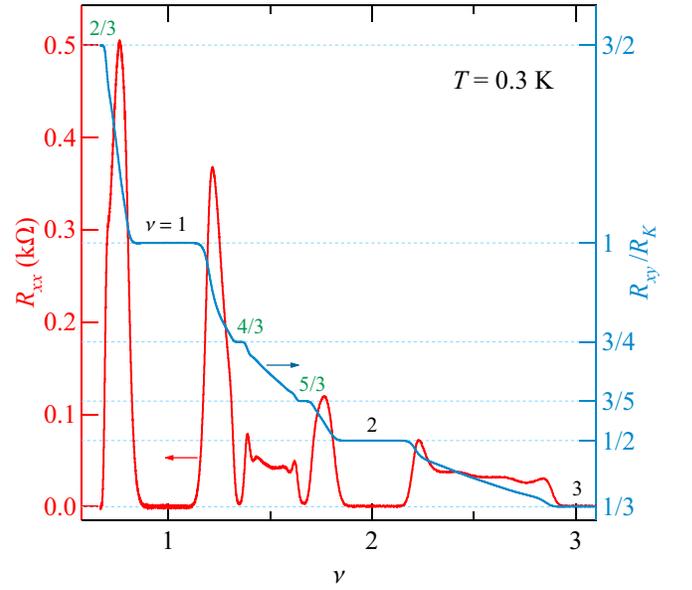}
\vspace{-0.25 in}
\caption{(Color online)
$\rxx$ (left axis) and $\rxy/R_K$ (right axis) as functions of the filling factor $\nu$.
Horizontal dotted lines are drawn at $\rxy/R_K = 1/3,1/2,3/5,3/4,1$ and 3/2.
Integers and fractions next to the traces mark filling factors.
}
\vspace{-0.25 in}
\label{fig2}
\end{figure}
To demonstrate the accuracy of the Hall quantization, we construct \rfig{fig2} showing $\rxx$ (left axis) and $\rxy/R_K$ (right axis), where $R_K \equiv h/e^2 \approx 25.812$ k$\Omega$ is the von Klitzing constant versus the filling factor.
Horizontal dotted lines, drawn at $\rxy/R_K = 1/\nu = 1/3,1/2,3/5,3/4,1$, and 3/2, accurately match corresponding plateaus in $\rxy$ at both integer and fractional $\nu$, attesting to excellent quality of Hall quantization.

\begin{figure}[t]
\includegraphics[width=\linewidth]{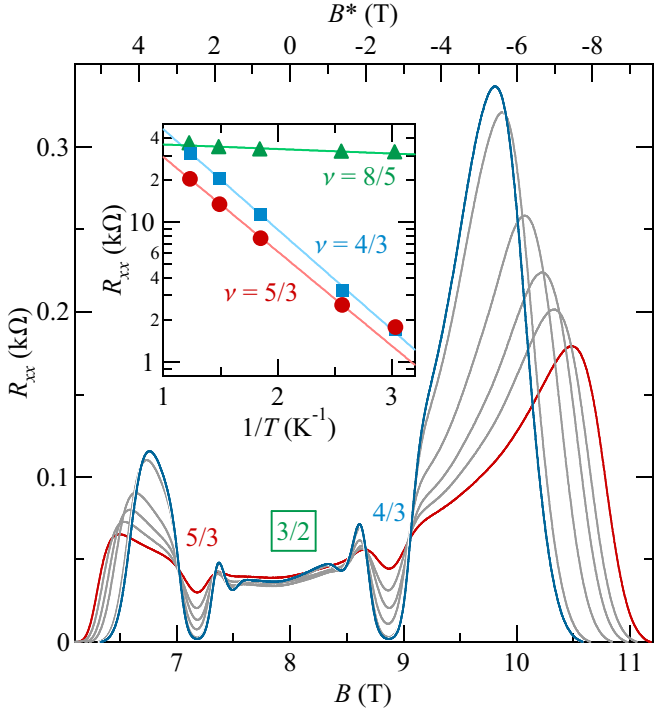}
\vspace{-0.2 in}
\caption{(Color online)
$\rxx$ as a function of $B$ (bottom axis) and $\bcf$ (top axis) in the $N=0$, spin-up Landau level at different temperatures from $T \approx 0.3$ K to $T \approx 1.1$ K. 
Inset shows $\rxx$ at $\nu = 5/3$ (circles), $\nu = 4/3$ (squares), and $\nu =8/5$ (triangles) versus $1/T$ on a log-linear scale.
The fits with $\rxx \sim \exp(-\Delta_\nu/2T)$ (solid lines), generate $\Delta_{5/3} = 3.1$ K and $\Delta_{4/3} = 3.3$ K.
}
\vspace{-0.25 in}
\label{fig3}
\end{figure}

Having confirmed quantization at filling factors $\nu = 5/3$ and $\nu = 4/3$, we now examine our data within the framework of composite fermions \citep{jain:1989,jain:2007}.
Around $\nu = 3/2$, composite fermions are formed by attaching two Dirac flux quanta ($2\fq = 2h/e$) to the empty states in the $N = 0$, spin-up Landau level.
Since the density of these states is given by $\ncf = (2/\nu - 1)p$, the composite fermions move in an effective magnetic field $\bcf = B - 2\fq\ncf = - 3(B - B_{3/2})$, where $B_{3/2}$ is the magnetic field at $\nu = 3/2$. 
Exactly at $\nu = 3/2$, $\bcf = 0$ and the composite fermions form a Fermi sea with the Fermi energy $\efcf$, determined by $\ncf$ and the composite effective mass $\mcf$.
Away from $\nu = 3/2$, composite fermions populate $\Lambda$-levels, separated by $\hbar\occf = \hbar e|\bcf|/\mcf$. 
As a result, the FQH states at $\nu = 2 - \nucf/(2\nucf \pm 1) = 5/3,8/5,...$ and $4/3,7/5,...$ can be viewed as integer quantum Hall states of composite fermions at $\nucf \equiv \ncf\fq/|\bcf|  = 1,2,...$ and $2,3,...$ respectively.

In a typical two-dimensional electron gas in GaAs, the Zeeman energy at $\nu = 3/2$ is of the order of 1 K, which is smaller than the Fermi energy of composite fermions.
As a result, the composite fermion system is often only partially spin-polarized which results in multiple crossings of spin-up and spin-down $\Lambda$-levels leading to suppression of select FQH states.
In our Ge sample, a rough estimate of the hole g-factor can be obtained by comparing the magnetic field onsets of quantum oscillations at even and odd filling factors.
Since the ratio of these onsets is close to two \citep{shi:2015a}, the ratio of the cyclotron energy to the spin splitting is about 3, yielding $g \approx (2m_e/m)/3 \approx 7.4$, where we have used the effective hole mass $m \approx 0.09\,m_e$ \cite{zudov:2014, morrison:2014}.
We can then estimate the Zeeman energy at $\nu  = 3/2$ as $E_z  = g \mu_B B_{3/2} \approx 40$ K.
Using the composite fermion effective mass obtained below, we find that this value is about three times larger than $\efcf = 2\pi\hbar^2p/3\mcf$. 
As a result, the FQH states are not affected by the spin degree of freedom and
the only parameter which determines the $\Lambda$-level spectrum is the composite fermion mass $\mcf$.
In what follows, we obtain $\mcf$ from the temperature dependence of magnetoresistance in the $N=0$, spin-up Landau level.

In \rfig{fig3} we present $\rxx$ as a function of $B$ (bottom axis) and $\bcf$ (top axis) at different temperatures from $T \approx 0.3$ K to $T \approx 1.1$ K. 
The data reveal that the resistances at the $\rxx$ minima at $\nu=5/3$ and $\nu = 4/3$ grow with increasing $T$ in a very similar fashion. 
This behavior suggests that the energy gaps at these filling factors are close to each other, indicating anticipated full spin polarization of both FQH states.
In the inset of \rfig{fig3} we present the $\rxx$ values at $\nu = 5/3$ (circles), $\nu = 4/3$ (squares), and $\nu =8/5$ (triangles) versus $1/T$ on a log-linear scale.
We find that both $\nu = 5/3$ and $\nu = 4/3$ data can be described reasonably well by exponential dependencies (solid lines), $\rxx \sim \exp(-\Delta_\nu/2T)$, with $\Delta_{5/3} = 3.1$ K and $\Delta_{4/3} = 3.3$ K, respectively.
In contrast, the $\rxx$ at $\nu = 8/5$ shows very weak dependence, indicating a vanishingly small gap ($\lesssim 0.1$ K).

Theoretically, the energy gap of FQH states at filling factors $\nu = n/3$, where $n$ is an integer, can be estimated as $\Delta  = \hbar \occf \approx 0.1 \ec$, where $\ec = e^2/4\pi\epsilon \epsilon_0 \ell = 41\sqrt{B[\rm T]}$ K is the Coulomb energy. 
Using this expression we obtain $\Delta_{5/3} \simeq \Delta_{4/3} \simeq 10$ K, which is considerably larger than experimental values.
It is well known, however, that finite thickness effects \citep{macdonald:1984,yoshioka:1984,dassarma:1986,yoshioka:1986} and Landau level mixing \citep{yoshioka:1984,sodemann:2013a} can reduce $\hbar \occf$ and that the experimental gap will be reduced even further due to the finite width of the $\Lambda$-levels $\gcf$.
One standard procedure to obtain $\mcf$ and $\gcf$ is to extract activation gaps for a series of FQH states and fit the data with $\Delta = e\hbar\bcf/\mcf - \gcf$ \citep{du:1993}.
While such a method cannot be reliably applied to our data, it can still give us a crude estimate. 
Using the two data points, $\Delta_{5/3} \approx 3.1$ K and $\Delta_{8/5} \approx 0.1$ K, we obtain $\mcf \approx 0.4\,m_e$ and $\gcf \approx 4.9$ K.

\begin{figure}[t]
\includegraphics[width=\linewidth]{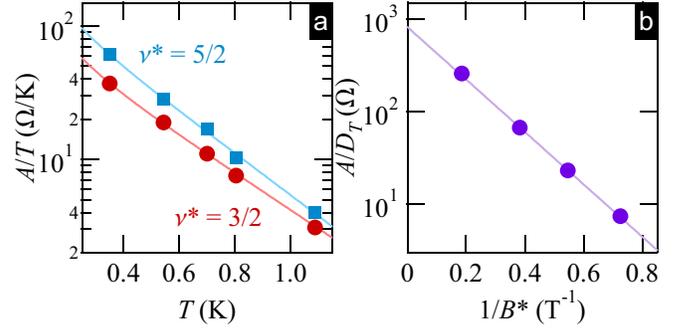}
\vspace{-0.2 in}
\caption{(Color online)
(a) Resistance oscillation amplitude $A$ normalized to temperature $T$ at $\nucf \approx  3/2$ (circles) and $\nucf \approx 5/2$ (squares) as a function of temperature.
Solid lines are fits with $A/T \sim 1/\sinh (2\pi^2 k_B T/\hbar\occf)$.
(b) Resistance oscillation amplitude $A$ normalized to $D_T = (2\pi^2 k_B T/\hbar\occf)/\sinh(2\pi^2 k_B T/\hbar\occf)$ as a function of $1/\bcf$.
Solid line is the fit to $A/D_T \sim \exp(-\pi/\tqcf\occf)$.
}
\vspace{-0.2 in}
\label{fig4}
\end{figure}

Another approach to obtain $\mcf$ is based on the Shubnikov-de Hass analysis using Lifshitz-Kosevich formula \citep{du:1994,leadley:1994}. 
More specifically, the $\rxx$ oscillation amplitude is expected to decay with increasing $T$ as 
\be
A \sim D_T = (2\pi^2 k_B T/\hbar\occf)/\sinh(2\pi^2 k_B T/\hbar\occf)\,,
\label{sdho}
\ee
where $k_B$ is the Boltzmann constant.
Such analysis can be most reliably performed at $B$ corresponding to the maxima of $\rxx$ located between the $\rxx$ minima at $\nu = 5/3$ ($\nucf = 1$) and $\nu = 8/5$ ($\nucf = 2$) , as well as at $\nu = 4/3$ ($\nucf = 2$) and $\nu = 7/5$ ($\nucf = 3$) .
These maxima occur near composite fermion filling factors $\nucf = 3/2$ and $\nucf = 5/2$, respectively, and the amplitudes can be estimated as, e.g. $A^{(\nucf = 3/2)} \approx \rxx^{(\nucf = 3/2)}/2 - [\rxx^{(\nucf = 1)}+\rxx^{(\nucf = 2)}]/4$.
Extracted in such a way amplitudes, normalized to $T$, are presented in \rfig{fig4}(a) as a function of $T$ on a log-linear scale for $\nucf = 3/2$ (circles) and $\nucf = 5/2$ (squares).
The fits to the data with $A/T \sim 1/\sinh (2\pi^2 k_B T/\hbar\occf)$ (solid lines) generate $\mcf \approx 0.42\, m_e$ and $\mcf \approx 0.45\, m_e$ for $\nucf = 3/2$ and $\nucf = 5/2$, respectively. 
By extrapolation, we can estimate $\mcf$ at $\nu = 5/3$ as $\mcf \approx 0.41\,m_e$ which corresponds to $\hbar\occf \approx 7.8$ K.
The $\Lambda$-level broadening parameter can be estimated as $\gcf = \hbar\occf - \Delta_{5/3} \approx 4.7$ K, which is close to the earlier estimate.
Finally, we find that at $\nu = 5/3$, $\hbar\occf \approx 0.07 \ec$, in reasonable agreement with the theoretical prediction, especially considering that it neglects the effects of finite thickness \citep{macdonald:1984,yoshioka:1984,dassarma:1986,yoshioka:1986} and Landau level mixing \citep{yoshioka:1984,sodemann:2013a}.

Another parameter which can be obtained from the Shubnikov-de Hass analysis is the quantum scattering time $\tqcf$ which contributes to the decay of the oscillation amplitude as the effective magnetic field is lowered.
The functional dependence of such a decay is given by $A \sim D_T\exp(-\pi/\occf\tqcf)$ and $\tqcf$ can be obtained from the Dingle plot analysis as illustrated in \rfig{fig4}(b).
Here, we plot the amplitude $A$, normalized by $D_T$ as a function of $1/\bcf$ on a log-linear scale.
The fit with $A/D_T \sim \exp(-\pi/\occf\tqcf)$ yields $\tqcf \approx 1.2$ ps.
With this value one can obtain the width of the $\Lambda$-levels $\gcf = \hbar\tqcf \approx 6.2$ K, which is reasonably close to earlier estimates.
The obtained value of $\tqcf$ is lower than the transport lifetime of composite fermions $\tau^\star\approx\,10$ ps, estimated from the resistivity at $\nu =3/2$. 
It is also lower than the quantum lifetime of 2D holes $\approx\,3$ ps, obtained from the Dingle analysis of microwave-induced resistance oscillations (measured in a different, but similar, sample) \citep{zudov:2014}.

In summary, we have observed and investigated the fractional quantum Hall effect in an ultra high-quality 2D hole gas in strained Ge quantum well.
The Hall resistance reveals plateaus at $\nu = 2/3, 4/3$ and $5/3$ which are quantized at $h/e^2\nu$.
From the analysis of the temperature dependence of the longitudinal resistance in the lowest, spin-up Landau level we determine the composite fermion mass of $\mcf \approx 0.4\,m_e$.
At $\nu =5/3$, this value corresponds to $\hbar\occf \approx 0.07 e^2/4\pi\epsilon_0\epsilon \ell$, in reasonable agreement with the theory. 
Due to large Land${\rm\acute{e}}$ $g$-factor, all observed fractions represent single-component FQH states which are fully spin-polarized and can be explained in terms of composite fermions with only orbital degree of freedom.

\begin{acknowledgments}
We thank A. MacDonald, S. Kivelson, and B. Shklovskii for discussions and G. Jones, S. Hannas, T. Murphy, and J. Park for technical assistance with experiments.
The work at the University of Minnesota was funded by the U.S. Department of Energy, Office of Science, Basic Energy Sciences under Award \# ER 46640-SC0002567.
Q.S. acknowledges Allen M. Goldman fellowship.
The work at the University of Warwick was supported by a EPSRC funded “Spintronic device physics in Si/Ge Heterostructures” EP/J003263/1 and “Platform Grant” EP/J001074/1 projects.
Experiments were performed at the National High Magnetic Field Laboratory, which is supported by NSF Cooperative Agreement No. DMR-0654118, by the State of Florida, and by the DOE.
\end{acknowledgments}



\end{document}